\documentclass{vldb}
\usepackage{graphicx}
\usepackage{listings}
\usepackage{url}

\usepackage{booktabs}

\usepackage[dvipsnames]{xcolor}
\usepackage{pgfplots}
\usepackage{pgfplotstable}
\usepackage{tikz}
\usetikzlibrary{positioning}
\usetikzlibrary{matrix}
\usetikzlibrary{shapes}
\usetikzlibrary{plotmarks}

\pgfplotsset{
    discard if/.style 2 args={
        x filter/.code={
            \edef\tempa{\thisrow{#1}}
            \edef\tempb{#2}
            \ifx\tempa\tempb
                
            \fi
        }
    },
    discard if not/.style 2 args={
        x filter/.code={
            \edef\tempa{\thisrow{#1}}
            \edef\tempb{#2}
            \ifx\tempa\tempb
            \else
                
            \fi
        }
    }
}
\newcounter{groupcount}
\pgfplotsset{
    draw group line/.style n args={5}{
        after end axis/.append code={
            \setcounter{groupcount}{0}
            \pgfplotstableforeachcolumnelement{#1}\of\datatable\as\cell{%
                \def\temp{#2}
                \ifx\temp\cell
                    \ifnum\thegroupcount=0
                        \stepcounter{groupcount}
                        \pgfplotstablegetelem{\pgfplotstablerow}{[index]0}\of\datatable
                        \coordinate [yshift=#4] (startgroup) at (axis cs:\pgfplotsretval,0);
                    \else
                        \pgfplotstablegetelem{\pgfplotstablerow}{[index]0}\of\datatable
                        \coordinate [yshift=#4] (endgroup) at (axis cs:\pgfplotsretval,0);
                    \fi
                \else
                    \ifnum\thegroupcount=1
                        \setcounter{groupcount}{0}
                        \draw [
                            shorten >=-#5,
                            shorten <=-#5
                        ] (startgroup) -- node [anchor=north] {#3} (endgroup);
                    \fi
                \fi
            }
            \ifnum\thegroupcount=1
                        \setcounter{groupcount}{0}
                        \draw [
                            shorten >=-#5,
                            shorten <=-#5
                        ] (startgroup) -- node [anchor=north] {#3} (endgroup);
            \fi
        }
    }
}

\usepgfplotslibrary{groupplots}

\usepackage{color}
\definecolor{olivegreen}{rgb}{0, 0.6, 0}
\newcommand{\jinho}[1]{}
\definecolor{antiquefuchsia}{rgb}{0.57, 0.36, 0.51}

\begin{document}
\title{System G Distributed Graph Database}

\numberofauthors{9}
\author{
\alignauthor
Gabriel Tanase\\
       \affaddr{IBM T.J. Watson Research Center}\\
       \affaddr{1101 Kitchawan Rd}\\
       \affaddr{Yorktown Heights, NY 10598}\\
       \email{igtanase@us.ibm.com}
% 2nd. author
\alignauthor
Toyotaro Suzumura\\
       \affaddr{IBM T.J. Watson Research Center}\\
       \affaddr{1101 Kitchawan Rd}\\
       \affaddr{Yorktown Heights, NY 10598}\\
       \email{tsuzumura@us.ibm.com}
       % 2nd. author
\alignauthor
Jinho Lee\\
       \affaddr{IBM T.J. Watson Research Center}\\
       \affaddr{11501 Burnet Rd}\\
       \affaddr{Austin, TX 78759}\\
       \email{leejinho@us.ibm.com}
       \and
% 3rd. author
\alignauthor Chun-Fu (Richard) Chen\\
       \affaddr{IBM T.J. Watson Research Center}\\
       \affaddr{1101 Kitchawan Rd}\\
       \affaddr{Yorktown Heights, NY 10598}\\
       \email{chenrich@us.ibm.com}
  % use '\and' if you need 'another row' of author names
% 4th. author
\alignauthor Jason Crawford\\
       \affaddr{IBM T.J. Watson Research Center}\\
       \affaddr{1101 Kitchawan Rd}\\
       \affaddr{Yorktown Heights, NY 10598}\\
       \email{ccjason@us.ibm.com}
% 5th. author
%\alignauthor Jie Lu\\
%       \affaddr{IBM T.J. Watson Research Center}\\
%       \affaddr{1101 Kitchawan Rd}\\
%       \affaddr{Yorktown Heights, NY 10598}\\
%       \email{jielu@us.ibm.com}
% 6th. author
\alignauthor Hiroki Kanezashi\\
       \affaddr{IBM T.J. Watson Research Center}\\
       \affaddr{1101 Kitchawan Rd}\\
       \affaddr{Yorktown Heights, NY 10598}\\
       \email{hirokik@us.ibm.com}
}
\additionalauthors{Additional authors:\\ 
Song~Zhang ({\texttt{zhangson@us.ibm.com}}) and \\
Warut D.Vijitbenjaronk({\texttt{Warut.Vijitbenjaronk@ibm.com}})\\
}
%\author{\IEEEauthorblockN{Gabriel Tanase, Toyotaro~Suzumura, Chun-Fu~(Richard)~Chen, Jason~Crawford, Jie~Lu, Hiroki~Kanezashi, \\ Ching-Yung~Lin, Danny~L.~Yeh, Song~Zhang}
%\IEEEauthorblockN{}
%\IEEEauthorblockA{IBM T.J. Watson Research Center\\
%Email: \{igtanase, tsuzumura, chenrich, ccjason, jielu, hirokik, chingyung, dlyeh, zhangson\}@us.ibm.com}
%\and
%\IEEEauthorblockN{The rest of the team}
%\IEEEauthorblockA{IBM TJ Watson Research Center\\
%Email: igtanase@us.ibm.com}
%}

% make the title area
\maketitle

\begin{abstract}
%abstract text goes here

Motivated by the need to extract knowledge and value from
interconnected data, graph analytics on big data is a very active area
of research in both industry and academia. To support graph analytics
efficiently a large number of in memory graph libraries, graph
processing systems and graph databases have emerged. Projects in each
of these categories focus on particular aspects such as static versus
dynamic graphs, off line versus on line processing, small versus large
graphs, etc.

While there has been much advance in graph processing in the past decades, 
there is still a need for a fast graph processing, using a cluster of machines with distributed storage.
In this paper, we discuss a novel distributed graph database called
System G designed for efficient graph data storage and processing on
modern computing architectures. In particular we describe a single
node graph database and a runtime and communication layer that allows
us to compose a distributed graph database from multiple single node
instances. 
From various industry requirements, we find that fast insertions and large volume concurrent queries are critical parts of the graph databases and we optimize our database for such features.
We experimentally show the efficiency of System G
for storing data and processing graph queries on state-of-the-art
platforms.

\end{abstract}

%\begin{IEEEkeywords}
%graph; database; distributed; analytics; dynamic;
%\end{IEEEkeywords}

\section{Introduction}
Data is generated at an increasing rate and researchers in industry
and academia are faced with novel challenges on how to extract
knowledge from it that subsequently can bring value to a community or
company. A first set of problems that need to be addressed when
working with big data is how to efficiently store and query them. In
this paper we focus on these challenges in the context of large
connected data sets that are modeled as graphs. Graphs are often used
to represent linked concepts like communities of people and things
they like, bank accounts and transactions, or cities and connecting
roads.  For all these examples vertices are in order of hundred of
millions and edges or relations are in the order of tens of
billions. For a property graph model \cite{neo4j2013} which is the
focus of our work, there can be an additional arbitrary number of
properties for each vertex and edge thus largely increasing the size
of data to be maintained and queried. 

The interest in graph databases has rapidly grown in the past decade with many discussions on multiple aspects including internal data structures and storage, query languages, programming models and analytic algorithms. As a result, a few open-source based graph database solutions have arisen~\cite{neo4j2013, janusgraph} and there are attempts to standardize public APIs and frameworks to query the graph data ~\cite{tinkerpop, gremlin}. Also, a number of programming models for fast processing of OLAP (\emph{On-Line Analytical Processing} work~\cite{malewicz2010pregel, powergraph, graphchi, extrav} have been proposed. However, there is not enough discussion on optimizing for performance in the distributed OLTP (\emph{On-Line Transaction Processing}) targeted graph databases and ways to design such databases are still under debate. Compared to OLAP-oriented approaches, OLTP systems need to support masive changes to the data usually with continuous, large volume insertions and numerous small-scale neighborhood searches. These performance requirements forces certain design choices and performance versus functionality trade-offs that need to be considered.

\sloppypar
In this work, we introduce the main components of the IBM System G graph database ecosystem:
a fast, ACID-compliant, single node graph database 
and a
distributed graph database that relaxes some of the ACID properties
in order to achieve high performance. The scenarios we are
currently targeting with our database are from financial world where
the graph is constantly updated with hundreds of thousand of edges per
second corresponding to transactions between various accounts while at
the same time performing thousands of queries per second to look for
special patterns in small-scale neighborhoods. 
Additionally, we found out that many users do not keep the graph database as their primary database engine. Instead, they often choose to migrate data from another type of database, often SQL, and run analytics on the graphs.
In those cases, building the graph may take several days to even weeks, dominating the time for the actual analytics phase. 
Therefore, the requirements for our graph database are: 1. Fast insertions of vertices and edges on the order of 100,000 per second. 2. Support hundred of thousands parallel queries, which are often small scale searches around a given vertex or edge.
We built SystemG because as far as we know, existing solutions could not provide enough performance for the requirements above, for some of our industry collaborators.

The contributions of our paper are a very fast single node graph
database implemented on top of a key value store and a distributed graph
database designed as a composition of a set of single node databases
referred to as shards. To support high throughput insertions and low
latency queries we introduce a novel runtime supporting thousands of
concurrent requests per second and a Remote Procedure Call (RPC)
based communication model to allow the database to move data
efficiently between various machines hosting different shards. On top
of our distributed design we introduce two novel techniques for adding
edges, techniques that reduces the number of exchanged messages
compared to a trivial implementation. We evaluate quantitatively
our design and compare the performance of our database with existing
solutions such as Neo4J~\cite{neo4j2013} and JanusGraph~\cite{janusgraph}.

The remainder of the paper is organized as follows. In Section
\ref{sec:related} we present related work, Section \ref{sec:single}
briefly introduces our single node graph database, Section
\ref{sec:dgraph} delves into our distributed graph database design
and organization. Section \ref{sec:bfs} describes a general approach
for writing graph queries on the distributed graph, Section
\ref{sec:methods} introduces two novel techniques for adding edges,
Section \ref{sec:consistency} describes the consistency model
supported by our current API, Section \ref{sec:experiments} include
our evaluation and comparison with other graph databases and Section
\ref{sec:conclusion} conclude our paper.

\section{Related Work}
\label{sec:related}
In this section we discuss related projects and we classify them into
four broad categories: graph datastructure libraries, graph
processing frameworks, where the emphasis is on the programming
models, graph databases, where the focus is on storage, and hardware based approaches.

{\bf Graph libraries}: Graph libraries for in-memory-only processing
have been available for a long time. For example Boost Graph Library
(BGL) \cite{boost} provides a generic graph library where users can
customize multiple aspects of the datastructure including directness,
in memory storage, and vertex and edge properties. This flexibility
gives users great power in customizing the datastructure for their
particular needs. Parallel Boost Graph Library \cite{pbgl},
STAPL \cite{stapl11} and Galois \cite{galois}, provide in memory
parallel graph datastructures. Pegasus~\cite{pegasus} implements graph algorithms as matrix multiplications, on top of MapReduce~\cite{mapreduce} framework, and GraphBLAS~\cite{graphblas} is another open-forum effort in a similar approach. 
All these projects provide generic
algorithms to access all vertices and edges, possibly in parallel,
without knowledge of the underlying in-memory storage
implementation. System G Native Store employs a similar design
philosophy with these libraries but it extends these works with
support for persistent storage and a flexible runtime for better work
scheduling.
% Should we reference a Native Store paper?
% There's a tilde before \cite several places below, but not above.  I don't know if that's intentional.

\jinho{flashgraph, prefedge}

{\bf OLAP Graph processing frameworks}: Pregel and
Giraph~\cite{malewicz2010pregel,giraph,scalegraph} employs a parallel programming
model called Bulk Synchronous Parallel (BSP) where the computation
consists of a sequence of iterations. In each iteration, the framework
invokes a user-defined function for each vertex in parallel. This
function usually reads messages sent to this vertex from last
iteration, sends messages to other vertices that will be processed at
the next iteration, and modifies the state of this vertex and its
outgoing edges. GraphLab~\cite{low2010graphlab} is a
parallel programming and computation framework targeted for sparse
data and iterative graph algorithms. PowerGraph~\cite{powergraph} extends the GraphLab with gather-apply-scatter model to optimize the distributed graph processing.
Pregel/Giraph and GraphLab are
good at processing sparse data with local dependencies using iterative
algorithms. PGX.D~\cite{pgxd} boosts performance of parallel graph processing by communication optimization and workload balancing. 
GraphX~\cite{graphx} is another Pregel based distributed framework that runs on Spark~\cite{spark}.

Another type of approach to reduce their network communication overhead is out-of-memory graph processing~\cite{graphchi, x-stream, gridgraph, llama, turbograph}.
They try to utilize the large storages (e.g., HDD, SSD). These frameworks are often focused on minimizing the number of storage accesses~\cite{graphchi, gridgraph}, utilizing sequential accesses~\cite{x-stream}, or maintain multiple outstanding I/O transactions to draw maximum bandwidth out of flash devices~\cite{flashgraph, turbograph}. There are also attempts to use mmap for efficient storage management~\cite{llama, mmap}. 
However, none of these are designed to answer ad hoc, small scale queries or
process graph with rich properties.

{\bf OLTP Graph processing frameworks}: TinkerPop~\cite{tinkerpop} is an open-source graph ecosystem
consisting of key interfaces/tools needed in the graph processing
space including the property graph model (Blueprints), data flow (Pipes),
graph traversal and manipulation (Gremlin), graph-object mapping
(Frames), graph algorithms (Furnace) and graph server
(Rexster and Gremlin Server). Interfaces defined by TinkerPop are becoming popular in the
graph community. As an example, Titan~\cite{titan} and JanusGraph~\cite{janusgraph} adheres to a lot of
APIs defined by TinkerPop and uses data stores such as HBase and
Cassandra as its scale-out persistent layer. TinkerPop focuses on
defining data exchange formats, protocols and APIs, rather than
offering a software with good performance.

{\bf Graph stores}: Neo4J~\cite{neo4j2013} provides a disk-based,
pointer-chasing graph storage model. It stores graph vertices and
edges in a denormalized, fixed-length structure and uses
pointer-chasing instead of index-based method to visit them. By this
means, it avoids index access and provides better graph traversal
performance than disk-based RDBMS implementations. 
McColl et.al \cite{graphdbsurvey2014} performed a thorough 
performance comparison of various open source graph databases. 

{\bf Hardware Based Approches}: There also have been attempts to take hardware into account for graph processing.
\cite{ceriani} introduces a computer architecture suitable for graph processing, tuning for irregular memory accesses. 
To exploit the parallelism, using GPUs is also a popular way. \cite{harish} presents CUDA implementations for graph processing algorithms, and \cite{bfs_gpu} optimizes BFS by combining top-down and bottom-up approaches.
There are custom hardware accelerators designed to draw maximum memory bandwidth and utilize the parallelism of graph processing~\cite{isca-acc, graphicionado}. ExtraV~\cite{extrav} is a near-data accelerator that uses coherent interface~\cite{capi} to increase the performance of out-of-memory processing. Tesseract~\cite{tesseract} places small computing units on the logic die of 3D stacked memories, and multiple blocks work together as an accelerator for graph processing. However, none of those approaches handle processing dynamic property graphs.

\section{Single Node Graph Database}
\label{sec:single}

The single node graph database implements a property graph model.
Each graph is identified by a user-specified name and consists
of vertices, edges, and the properties (i.e. attributes) associated
with vertices and edges.  Each vertex is identified by a unique
external vertex identifier specified by the user and an automatically
generated unique internal vertex identifier.  Each edge is identified
by the vertex IDs of its source and target vertices and an
automatically generated unique edge ID.  Multiple edges between the
same pair of vertices are allowed. 
Each vertex or edge is associated with a label string which can be
used to categorize vertices/edges and facilitate efficient traversal
(e.g. only traverse edges of a specific label).

A rich set of graph APIs is provided to support all fundamental graph
operations, including graph creation/deletion, data ingestion (add
vertex/edge one at a time or via batch loading of files in CSV
format), graph update (delete vertex/edge, set/update/delete
vertex/edge properties), graph traversal (iterate through vertices and
edges of each vertex), data retrieval (get vertex/edge properties),
graph search (find vertex/edge by ID, build/search property index).
  
  Internally, vertex-centric representations are used to store vertices
and edges, along with the maps for vertex and edge properties.  For
the current version evaluated in this paper, LMDB (Lightning Memory-Mapped Database)~\cite{lmdb}, 
a high-performance key-value store is used to store the above
representations both in memory and on disk. In the rest of the section, the implementations of graph components using key-value stores will be explained in detail.

\subsection{Vertices}
\label{subsec:vertices}
Information about the existence of individual vertices is maintained by two key-value store databases: \emph{ex2i} (external\_id-to-internal\_id) and \emph{i2ex} (internal\_id\_to\_external\_id).
The \emph{ex2i} uses the external identifier as the key and internal id as the value, and vice versa for \emph{i2ex}.  
As mentioned above, we assume the user always accesses a vertex using its external identifier. A unique internal id exists for each external ids, and those internal ids are used in reference to other components (i.e., edges and properties). The reverse-mapping database, \emph{i2ex}, is also maintained to convert the internal ids back to the external id and make this information available to the user.

\subsection{Edges}
\label{subsec:edges}
Edges are managed by the \emph{vi2e} key-value database. In \emph{vi2e}, the source vertex internal identifier is used as the key to locate edges outbound from the vertex.
The value stored for each edge includes the internal id of the target vertex, label id, and the edge id. 
The edge id is unique, and is used to distinguish between the edges with the same source and target vertices. 
Also, it is used to index the properties of edges.
In the case where the graph is directed, SystemG maintains another key-value store database for tracking inbound edges from a vertex. The structure is exactly the same as \emph{vi2e}, except that the destination vertex id is used as the key. 
The same edge identifier is used between the two edge databases (i.e., inbound and outbound) to mark that they belong to the same edge instance.

\subsection{Properties}
\label{subsec:properties}
  
  The property set of a 
vertex/edge is essentially a list of key-value pairs where each key is
a property name and the value associated with the key is the value of
the corresponding property for this vertex/edge.
Property values can
be strings, numbers (integer, float, double), vector of numbers, or
composite values consisting of strings and numbers.  Multiple values
for a single property, and properties (e.g. meta data) of properties
are supported for compliance with Apache TinkerPop 3.

There are two property key-value store databases, \emph{vid2pkv} and \emph{eid2pkv}, each for vertex properties and edge properties, respectively.
 As in the vertex ids, the property names are mapped into a property ids and the property ids are used for those two databases.
In those databases, the vertex id or the edge id is used as the key, and the property id and the property value are stored as the value.
Since LMDB allows entries with duplicate keys, and stores values in a sorted order, the correct property can be found by looking for the matching property id. 
Alternatively, one could concatenate vertex id/edge id with the property id and use the concatenated data as the key. In that way, the key will always be unique and no matching for property ids is required. 
However, we found out that using an integer (the vertex id) as the key gives a hint for internal optimizations in the key-value store, and makes it often faster than the alternative method avoiding the property id matches, unless the number of properties is large. 
We provide the trade-off as a configuration option for the database, and use the former method in the remaining of the paper.
\jinho{maybe pname2pid}

\subsection{ACID Transactions}
\label{subsec:acid}
The single node graph database is fully ACID compliant by exploiting the feature of the underlying key value store used which in our case is
LMDB. The graph database exposes a transactional API where users can
start a read/write or read only transaction. The LMDB uses copy-on-write 
technique for transactions and under this model a read/write
transaction will create new memory pages with the data being modified
and these pages will be made visible to the other threads when the
transaction commits.  While multiple threads or processes can start read/write
transactions, they will be serialized by the database with only one
being active at any time. Read only transactions are not restricted
from accessing the database at any time and they can all proceed in
parallel.

\section{Distributed Graph Database}
\label{sec:dgraph}

\begin{figure}[htp]
\centering
\includegraphics[width=0.48\textwidth]{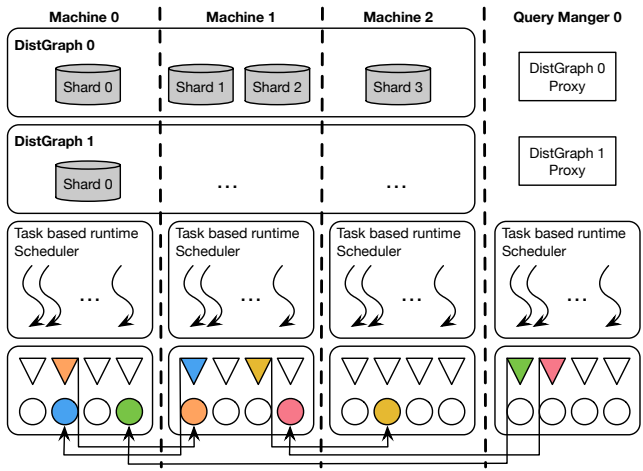}
\caption{Distributed draph database organization.}
\label{fig:dgraph}
\end{figure}

The distributed  graph database  is a  composition of  a fixed  set of
single  node  graph databases  called  shards.  The distributed  graph
maintains a list  of computation nodes, the mapping  of shards to
nodes and  implements an  API such  that callers  see only  one database
instance  and not  a  collection of  distributed  services. Thus  upon
instantiating a distributed  graph, a naive user will  have access to
the  same  interface  as  with   the  single node graph  database  and
the location of various vertices and edges is abstracted away
by the distributed graph API.

The graph distributes its vertices by default based on a hash function
applied to the external vertex identifier.  An edge will be located
with its source vertex by default.  Thus a typical distributed graph
method will perform as its first step the computation to determine the
shard where a particular vertex or edge is or will be located.
Subsequently the method invocation will be forwarded using Remote
Procedure Calls (RPC) to that shard to complete the method execution.

In Figure \ref{fig:dgraph} we show a distributed system running two
distributed graph instances. In general a machine can host an
arbitrary number of shards, where each shard runs as a distinct
process. Each shard is uniquely mapped to a host and port as specified
by a hostfile created during the cluster configuration. The cluster
depicted has three physical machines hosting the distributed graph
system: Machine 0, 1, and 2. Within one machine we have a number of
shards belonging to possibly multiple graph databases. All the shards
of a database are managed by a distributed graph object responsible
for address resolution and method forwarding.

Data in a distributed graph database can be accessed by any of the
nodes of the database using RPC. Additionally, in our model we
introduce the notion of QueryManager (QM) as a separate process
running on its own machine with its own hardware resources. The QM
will instantiate a distributed graph object in a proxy mode
in which case it will not own any data and any request received by
the QM will be forwarded to appropriate shards.

%In the following section we introduce our main communication
%primitives that are used to exchange data between the nodes of the
%database

\subsection{Messaging Layer}
The graph database will run in a Single Program Multiple Data (SPMD)
fashion similar to MPI. The binary corresponding to an application
(e.g., graph database) will be executed on multiple machines and each
process will have its own identity and know how many other processes
are part of the overall computation. After an application starts it
can access local memory and local storage.  When remote data needs to
be processed, communication will be employed.  Our distributed graph
system uses Remote Procedure Call (RPC) as its core communication
abstraction.  The RPC is abstracted on top of a native communication
library like sockets, MPI, PAMI or GASNet, inheriting advantages and
disadvantages of the underlying layers.  The RPC  provides
to the distributed system developers a high level abstraction that
improves productivity and portability of the system.

\subsection {Runtime}
In general each individual process will receive RPC requests from
multiple sources. In order to provide a high throughput (executed
RPCs per second) we employ a multithreaded task based runtime. Within
our system, each RPC invocation received from the network is
encapsulated within a task and placed into the runtime scheduler for
execution. The runtime scheduler maintains a pool of worker threads and
dispatches individual tasks to individual threads. The scheduler also
allows for work stealing to keep the load balanced. The same runtime is
also used within our framework to execute parallel computations within
one SPMD node.

%The runtime scheduler maintains a pool of threads, each with its own
%queue of tasks to be executed. When a new task is created the
%scheduler will place it into one of the queues in Round Robin
%fashion. Idle threads will also periodically check other threads
%queues and steal tasks if available.
%
When an RPC request is received on one of the incoming communication
channels, the messaging layer will receive the message and
quickly queue a task for execution by the worker threads. 
%This way the polling
%thread can extract RPC requests and posts them for execution very
%fast. 
The number of RPC's that can be executed concurrently
will be proportional to the number of worker threads used by the
scheduler. Individual RPC can invoke additional RPCs as part of their
body thus allowing us to implement certain functionality in a distributed
asynchronous fashion.
% ccjason: I modified much of the paragraph above.  I'm not sure exactly what we want to say here.  Hopefully I've gotten it right

\subsection{Query Manager}
\sloppypar
The System G architecture includes two
categories of clients: {\em query managers} and {\em regular clients}. 
The query manager is a
database client that is usually deployed on a powerful node.
The query
manager will open a communication channel with each individual
shard of the distributed database and afterwards is capable of
posting RPCs to any of the nodes of the database to perform
various graph operations. Regular clients will not connect to the
distributed database directly but rather they will connect to the query
manager and the query manager will execute the query on behalf of the client.

The extra level of indirection has the following advantages. Firstly it
allows the query manager to control the load on the database. The query manager may
decide to briefly delay the processing of some requests rather than overload the database.
Secondly it is often the case that
a query on a graph database is a complex computation like a breadth
first search (BFS) or finding a path between a source and a
target. For these queries the query manager may end up accessing all
shards of the database multiple times. Thus the query manager can
maintain all this partial state while performing the query and it
returns to the client when the final answer is available.

%It is also
%possible that the query manager can perform additional number of
%optimizations like some information cashing to further optimize
%response times for advanced read queries.
%
%The query manager when creating an access point to a graph database
%will instantiate the same distributed graph class as all the other
%processes of the database except that all of its data will be remote
%and accessed using RPC.
%
There can be more
than one query manager per system but usually not more than tens.
On the other hand, there can be hundreds of clients communicating
with the query managers over a network protocol.
Currently a query manager provides multiple interfaces, such as REST, nanomsg~\cite{nanomsg} or native C++.
For REST interface, a query manager can start an HTTP server and
accept REST queries from clients that are subsequently mapped into
graph operations. The REST requests can be issued from a browser,
JavaScript, Java or Python program.

\renewcommand{\ttdefault}{pcr}
\lstset{%
  %basicstyle=\setstretch{0.89}\ttfamily, %\footnotesize,
  basicstyle=\ttfamily\small, %\footnotesize,
  numbers=left, firstnumber=1,
  xleftmargin=12mm, framexleftmargin=5mm,
  emph={BFS, graph, Class, inherit, do, while, wait, until, for, return, Procedure, add_edge}, emphstyle=\textbf\textit\bfseries,
escapeinside={\%}{)}  }
    
\begin{figure*}[h]
\centering
\begin{tabular}{c}

%\footnotesize
\begin{lstlisting}
BFS (distributed_graph* dg, id_type vid, int max_depth){
  partitioned_frontier frontier;
  frontier.insert(vid);
  d=0;
  while(d < depth && frontier.size() > 0) {
    for(idx_t nid=0;nid<frontier.locations();++nid){
      vector<id_type>& curr = frontier->get_verts(nid);
      if(curr.size() > 0){
	dg->get_all_edges_async(curr, new_verts[nid]);
      }
    }
    //in a second phase we wait for data to come back;
    for(idx_t nid=0;nid < frontier->locations();++nid){
      vector<idx_t>& curr = frontier->get_verts(nid);
      if(curr.size() > 0){
        while(new_verts[nid].second == 0) {
	  dg->poll(); //wait for vertices to come back
	}
	//process the newly arrived vertices
	for(idx_t j=0;j < new_verts[nid].first.size();++j){
          ...
 }}}}}
\end{lstlisting} \\

\end{tabular}
\caption{Fixed depth BFS.}
\label{fig:bfs}

\end{figure*}

\subsection{Firehose}
\label{sec:firehose}

In many large scale practical graph applications, the graph database
needs to process a large number of vertex or edge creation requests in
addition to read-only query requests.  
In this case, directing all ingests to the query manager often overloads the query manager and creates a bottleneck.
To address these applications,
we introduce the Firehose module, an extension for optimizing the
ingestion of data. The single node graph database that we use to
implement the distributed graph database is optimized for a single
writer, multiple readers scenario. For this reason in our design each
process running a shard of the database creates an additional thread
that is in charge of read-write transactions.  For these applications
the Firehose module will be in charge of requests requiring write
access, possibly in a batched mode.
By bypassing the query manager and directly talking to the dedicated thread in the shards, much higher bandwidth for vertex/edge ingestion can be achieved (see Section~\ref{sec:experiments}).
while query manager as introduced
in this section, will be mainly in charge of read-only queries. 
%\jinho{a little more detail, maybe. +kafka?}

\section{Graph queries and analytics}
\label {sec:bfs}

The query manager will handle all basic queries like add/delete/get
vertex, edge, property. Additionally the query manager will implement
graph specific queries like various traversals. In this section we
describe how queries for the distributed graph can be internally implemented.  As
a simple example let's consider a simplified breadth first search
where we traverse only a given number of levels deep. The relevant source
code is included in Figure \ref{fig:bfs}.

As parameters, the analytic takes a reference to the distributed graph instance, the
starting vertex and the number of levels to traverse (line 1). First it instantiates a
partitioned frontier which will maintain a list of vertex identifiers
grouped by the shard to which they belong. We start by adding the
initial starting vertex to it (Lines 2-3). Next we start an iterative
process where we compute the next BFS frontier based on the current
frontier. For this we extract from the frontier vertices that all live
in a certain shard and we post an asynchronous request to that shard
to collect and retrieve all neighbors of this set of source vertices
(lines 7-10). This is performed using the graph method {\tt
  get\_all\_edges\_async()} which internally will use the RPC
mechanism previously described. This method runs asynchronously, allowing the calling function
to submit multiple requests before polling for results (line 17).
It is possible that by the time the
last method invocation is finished, some of the results are available, thus this
flexible RPC mechanisms allows us to overlap communication with data
retrieval. 
%In a second phase the program waits for data from a particular shard to arrive
%and next processes the received data preparing the next wave of the
%BFS. 

 We also implement the Pregel runtime so that developers can write
their graph analytic algorithms in a vertex-centric manner. By having
the Pregel runtime, the computation is also performed in a parallel
and distributed manner. Thus such a tightly coupled integration of
distributed databases and the distributed computing framework is also
one of the novel features. More detailed information on the Pregel
runtime will be described in a separate paper.
% due to space limitation. 

%Additional analytics can be implemented in a similar execution
%model with the algorithm described in this section.

\begin{figure}[htp]
\centering
\includegraphics[width=0.48\textwidth]{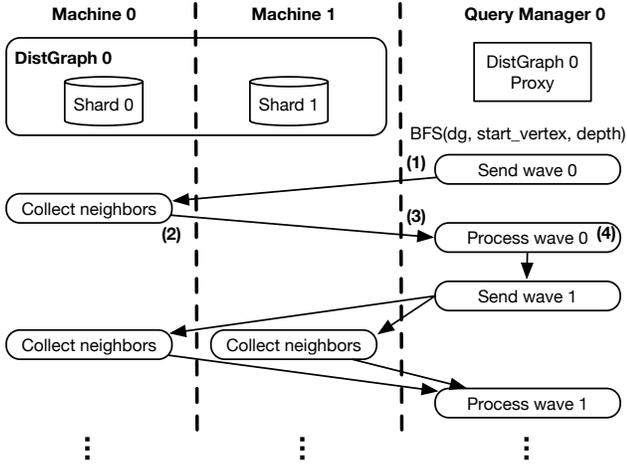}
\caption{Computation model performance estimates.}
\label{fig:model}
\end{figure}

%In the future we
%plan to start exploiting distributed asynchronous algorithm to perform
%various analytics.
%
{\bf Algorithm cost analysis}:  At a very
high level the BFS algorithm introduced in the previous section
consists of a set of rounds as shown in Figure \ref{fig:model}.  Each
round spawns a set of asynchronous requests (1) followed by remote
execution of the requests (2) followed by the results being sent back
(3) and the postprocessing step where the new frontier is assembled
(4). 

For a given architecture (compute nodes, storage network) one can
measure individual performance metrics that can be used to estimate the
overall cost of the computation. Thus the total time for one iteration
of the BFS algorithm can be estimated as
$T_{step}=T_{(1)}+T_{(2)}+T_{(3)}+T_{(4)}$. The total time will be
$T=\sum_{i=0}^{depth} T_{step i}$. To approximate $T_{(1)}$ we need to
have an estimate of the latency ($Lat$) and bandwidth ($Bw$) of the
network used. The values for this parameter is usually in order of a
few hundreds microseconds for a gigabit type network. The time to read
the neighborhood of a set of vertices ($T_{(2)}$) depends on the performance of
the key value store utilized which in turn depends on the computation
node and more specifically the storage used. This time is often in the
order of one millisecond range. Returning the neighbors back from the
network  ($T_{(3)}$) is similar to $T_{(1)}$ except the amount of data returned
may be different that step (1). In general this can vary from few
hundred microseconds to milliseconds range. The time for step (4) is
the time for processing the list of neighbors returned by RPCs. This
is often in memory computation and it is in the order of nanoseconds
to a few microseconds. Thus the overall query time is dominated by
the disk access time to read the neighborhood of a vertex.

\section{Vertex and edge management}
\label{sec:methods}

In the distributed graph database presented here each vertex and edge
is uniquely identified by an internal vertex and edge identifier
respectively. In this section we discuss how identifiers are generated
and managed while adding items to the database. Edges (outgoing and
incoming) are stored as tuples of such identifiers to save storage and
improve the data lookup performance. Additionally vertex and edge
properties are stored as key, value pairs using the vertex or edge
identifiers as keys as explained in section~\ref{subsec:properties}.

\sloppypar{{\bf Internal Vertex identifiers} Each vertex has a unique numeric
internal identifier. This is allocated when the vertex is created and it won't
be reused for any other vertex in the database. In a single node graph
database producing an unique id is done by incrementing a variable
each time a vertex is added. We will refer to this variable as
MAX\_VID. For a distributed graph database we ensure a unique vertex identifier
by using the following protocol. First a vertex is uniquely associated
with a shard by using either a default hash function or an arbitrary
placement function provided by the user.  When adding a vertex to a shard
a vertex identifier is generated by incrementing the shard-local
MAX\_VID variable and the overall global identifier of the vertex
becomes the following triplet $\{LabelId, ShardId,
LocalVertexIdentifier\}$}. 
Thus, when locating a vertex with certain internal id, 
its shard number can be easily found by looking at the few bits within the id that represents shards.

{\bf Edge Identifiers} Each edge has associated a unique edge
identifier. A major challenge for our design comes from the fact that
we allow multiple edges between the same two vertices and because we store
both incoming and outgoing edges. Assuming we have two vertices A
and B and we add an edge $\{A,B\}$ followed by another edge from
$\{A,B\}$. Using a unique edge identifier allows us to distinguish
between the two edges: $\{A,B,eid1\}$ and
$\{A,B,eid2\}$. For undirected graphs or graphs where we track
incoming edges we store two edge tuples in our database. For
example, for the edge $\{A,B\}$ we store one outgoing edge
$\{A,B,eid1\}$ and one incoming edge $\{B,A,eid1\}$. Both edges will
know they are part of the same edge because they share the same unique
identifier.

For a single node graph database an edge identifier can be easily
generated by incrementing a MAX\_EID variable. For the
distributed database the number of actions to be performed when adding
an edge increases due to the fact that the source and the target may
live in two different shards on two different machines. Assuming a
vertex A is mapped to shard1 and a vertex B is mapped to shard2, for the edge $\{A,B\}$ we store one outgoing edge
$\{A,B,eid1\}$ on shard1 and one incoming edge $\{B,A,eid1\}$ on
shard2. The edge identifier can be generated by shard1 and
communicated to shard2 together with the rest of the arguments when
adding the incoming edge. It is also valid to generate the id in
shard2 and communicate to shard1. In either case the shard identifier
also must be embedded in the most significant bits of the edge
identifier to make it globally unique.
\\

\subsection{Efficient edge addition for a distributed database}
A very common operation for graph databases is adding an edge between a
source and a target vertex without adding vertices in a prior step. For
example {\tt add\_edge(A, Knows, B)}. This turns out
to be a complex operation as shown in Figure \ref{fig:ae}. First the
operation will add two vertices A, and B if they don't
exist already (Figure \ref{fig:ae}, Lines 2 and 3), create a label
``Knows'' if it doesn't exist already (Figure \ref{fig:ae}, Line 4),
add the outgoing edge for A (Figure \ref{fig:ae}, Line 5) and
add the incoming edge to B (Figure \ref{fig:ae}, Line
6). Note that each of these invocations produces vertex and edge ids
that are subsequently used creating data dependencies between the five
steps of the method.

\begin{figure}[htb]
{
\scriptsize
\begin{lstlisting}
add_edge(A, Knows, B)
 VIDS=check_or_create(A)
 VIDT=check_or_create(B)
 LID=check_or_create_lid(Knows)
 EID=add_outgoing_edge(VIDS,VIDT,LID)
 add_incoming_edge(VIDT,VIDS,EID,LID)
\end{lstlisting}
}
\caption{Add edge steps.}
\label{fig:ae}
\end{figure}

\subsection{Basic algorithm} 
\label{sec:sync}

A straightforward approach to implement the steps depicted in Figure
\ref{fig:ae} is to execute the code on a client or query manager
node and execute all 5 steps synchronously one after another. Thus for
each of the steps 2 to 6 except step 4 we will have two messages exchanged over the
network: one to invoke remotely the operation and one to return
results used in the subsequent steps. Step 4 does not incur network traffic since checking for label ids is often performed locally by the query manager using caching. 
Thus there will be a total of at
least seven messages exchanged (the last step doesn't have to return
anything). If confirmation of the final step is required then this
approach will take a total of eight steps. 
To reduce the number of steps, SystemG optionally stores an in-memory cache of vertex list, which keeps the recently used external id to internal id mapping. If the cache hits, there is no need for message round-trip to the corresponding shard. For the rest of the paper, we assume that this cache is deployed.

\subsection{Asynchronous algorithm} 
\label{sec:async}
A first improvement we propose in this paper is to use asynchronous
RPC mechanism that our runtime natively supports. For this approach we
first forward the add edge method to the node where the destination is
allocated (DEST\_SHARD). On this shard the destination vertex is found
or created and its id (VIDT) is forwarded with the rest of the
arguments to the machine where the source is located
(SOURCE\_SHARD). Here the source vertex will be located or created
(VIDS), the outgoing edge will be created $\{VIDS,VIDT,LID,EID\}$ and
finally forward the invocation back to DEST\_SHARD to add the incoming
edge using the edge id previously generated. Thus we reduce the number
of communication steps from seven down to three. A fourth step can be
optionally employed if a confirmation of the method termination is
required on the client initiating the operation.

\subsection{Batched edge and vertex addition}
%\label{sec:firehose}

It is often the case that edges and vertices are added to the database
at very high rates and it is acceptable by the user's application
that the vertices or edges are added in a delayed, batched fashion. In this
section we describe a novel mechanism for adding items using
batches. We introduced in Section \ref{sec:firehose} the notion of
Firehose for optimizing fast insert rate operations and the batching
mechanism presented in this section is implemented as part of the
Firehose. Let's assume a set of edges are to be added to the database
using the semantic described in Figure \ref{fig:ae}. The Firehose will
collect a batch of them of size N and perform the following processing:

\begin{enumerate}
\item Create 2$\times$P queues where P is the number of shards. For each shard there will be one outgoing and one incoming edges queue.
\item For each add edge request place one entry in the outgoing edges queue corresponding to the shard where the source vertex of the edge is allocated. Similarly we place an entry in the incoming edges queue corresponding to the shard where the target vertex is allocated.
\item Instead of protecting the queues with locks, we create another set of queues to do double-buffering and avoid the problem of simultaneous reader-writer problem. 
\item For each pair of queues for each shard we collect the set of vertices to be added to the shard and we send one bulk request to the shard to add the vertices. This step can be done in parallel for all pairs of queues and their corresponding shard. The request will return the vertex identifiers for all newly added vertices. It will also reserve an edge id range on the shard and the edge id range is also returned to the Firehose.

\item The mappings from external vertex id to internal vertex ids are inserted into a map datastructure (cached) as subsequent steps will look for this mapping.

\item Based on all vertex identifiers returned and edge ranges reserved the Firehose will prepare the final tuples corresponding to the edges to be added. The edge tuples containing only internal ids will be sent to the database shards to be inserted. This insertion also happens in parallel for all shards.
\item Optionally, the mapping from external vertex id to internal vertex id can be cached on the Firehose such that to minimize the number of vertices information sent to the shards in step 3.
\end{enumerate}

This novel approach for performing batched edge addition provides the
highest amount of parallelism and the lowest number of messages
exchanged compared to the other two methods previously introduced. For
a given batch of N method invocations, the basic algorithm will
perform 7$\times$N communication messages, synchronizing for each step. The
asynchronous algorithm performs 3$\times$N messages if the invoking thread
doesn't require confirmation termination or 4$\times$N messages if
confirmation is required. The batched approached will exchange four
larger granularity messages per shard for the whole set of N
invocations for a total of P$\times$4 messages. Usually P will me much
smaller than N. While the batched method sends much fewer messages,
there is more data per message sent. However most networks perform
better when data is aggregated in bigger chunks.

\section{Fault tolerance, transactions and consistency model}
\label{sec:consistency}

When designing a distributed graph database one faces additional
challenges that one does not face designing a single node graph
database. In this section we discuss the ACID properties, consistency
model and fault tolerance guarantees. 

As discussed in Section \ref{sec:single} the single node database is
fully ACID compliant. 
While the single node graph database is fully ACID, our distributed 
graph database implementation is not. Currently we don't support
transactions for arbitrary sets of operations acting on multiple
shards. We are only ACID within one shard. In our model adding one
vertex and its associated properties is atomic and durable. However
adding two or more vertices is considered to be multiple transactions
with one for each vertex added. When adding edges, that is also a
composed operation consisting of one transaction to add the outgoing
edge and another transaction to add the incoming edge. These
transactions may happen on two different shards.

%The complex add edge
%operation as described in Figure \ref{fig:ae} will cause a variable
%number of transactions depending on the scenario used. For the SYNC
%implementation (Section \ref{sec:sync}) there will be one transaction
%to add the source vertex, a second to add the target vertex, a third
%to add the outgoing edge and possible properties and a fourth to add
%the incoming edge. For the ASYNC case (Section \ref{sec:sync}) we have
%one transaction to add the target vertex, a second to add the source
%vertex, the outgoing edge and its properties, and a third transaction
%to add the incoming edge. 
%
%For the Firehose method of ingesting edges, we have one transaction to
%bulk add all vertices for each shard for a total of P transactions
%where P is the number of shards. This transaction also reserves NE
%edge identifiers that will be used in a subsequent step. A second
%transaction will add all outgoing and incoming edges and their
%properties, for each shard for a total of P transactions. So for each
%batch there are 2$\times$P transactions.
%
Deleting an edge can also involve two transactions. Deleting a vertex
is the most complex operation on the distributed graph database as it
causes one transaction where the vertex lives to remove the vertex,
its properties and its outgoing edges and their properties, followed
by NI transaction on shards hosting edges pointing to this vertex,
where NI is the number of incoming edges.

\subsection{Memory Consistency}

For distributed algorithm writers the fact that we are not fully ACID
requires certain caution when accessing the data. For example when
looking for an edge one should not assume that both outgoing and
incoming ends are present.  For a transient interval of time
the fact that the outgoing end exists, doesn't imply that the
incoming end will be found. Additionally, one must retrieve a vertex
and its adjacency as one operation to ensure
atomicity. For example if two operations are used, the second one may
fail to find the vertex if an intervening delete operation occurred.

Unlike a single node graph database, a BFS traversal on the
distributed graph is not one transaction. For BFS accessing the 
neighborhood of each vertex
is a transaction, but it is possible that edges and vertices will be
added/deleted between collecting adjacencies for multiple
vertices. For the practical situation for which we are using the
current solution, data is mainly added to the database, so this
modality of querying will return all the adjacency when the query
started plus some eventual new vertices and edges added since the
query started. The fact that vertices/edges can be deleted while the traversal is in
progress requires the algorithm writer to not assume that if a
vertex was found during one call, it will be found on the next
call. 

While this programming model sounds complex it does allow the user to
write fast queries. While providing full transaction support for
distributed graph is on our agenda, currently we suspect this will come
at a great performance cost, requiring the database to potentially
lock most of the shards for read and write transactions.

\begin{table}[b!] 
\small
\centering
\begin{tabular}{llc}
\toprule
\multicolumn{2}{l}{Resource} & Value \\
\midrule
\multicolumn{2}{l}{Server Product}         & IBM S824L\\
\multicolumn{2}{l}{NUMA nodes} & 4\\
\multicolumn{2}{l}{Number of cores} & 24 \\
\multicolumn{3}{l}{Core:} \\
 & Frequency                       & 3.3GHz \\
 & SMT                    & 8 \\
 & TLB         & 2048 entries\\
\multicolumn{3}{l}{Cache:} \\
 & L1  & Private 64KB \\
 & L2      & Private 512KB per core \\
 & L3          & Shared 8MB per core \\
 & L4          & Shared 16MB per DIMM  \\

\multicolumn{3}{l}{Main Memory:} \\
%%%%% & Number of channels 		     & 1 \\
 & Technology             & DDR3 \\
 & Capacity       & 2TB \\
 & Bandwidth                 & 460GB/s \\
 & Page size                 & 4KB \\
 & Cache line size & 128B \\
 
\multicolumn{2}{l}{Storage} & IBM Flashsystem 840\\
% & DRAM refresh interval ($tREFI$)   & 7.8$\micro$s \\
\bottomrule
\end{tabular}
\vspace{-1mm}
\caption{System parameters of the server machines}
\vspace{-3mm}
\label{table:parameters}
\end{table}

\subsection{Fault Tolerance} In the current version of the database there is no fault tolerance
implemented. In our experience the graph database is often not the
main data repository but it is often used as an accelerator for graph
queries and it runs along side the main data repository. We support the use of Apache Kafka~\cite{kafka} for fault-tolerance communications in firehose ingestions. However, we found Kafka to be not a good fit for normal queries. Kafka is highly optimized for bandwidth and tradeoffs the latency. 
Also, Kafka does not naturally support request-reply style communication, making it hard to use for queries that have return values.
We do have
work in progress to implement a fault tolerance protocol by
providing replicas for the shards.

\section{Experimental Results}
\jinho{redoing the experiments for two reasons\\
1. show scalability for with four worker machines. current (two) seems too small. \\
2. show that our system scales well by also showing the scalability of neo4j and titan (Janusgraph)}

%for experimental section

\begin{figure*}[t]
\begin{minipage}[b]{0.48\textwidth}
%\centerline{\includegraphics[scale=0.6]{experiments/add_vertex.eps}}

\pgfplotsset{
% override style for non-boxed plots
    % which is the case for both sub-plots
    every non boxed x axis/.style={} 
}

\begin{tikzpicture}
    \begin{axis}[
            ybar,
            bar width=.5cm,
            width=\textwidth,
            height=12em,
            %legend style={at={(0.5,1)},
            %    anchor=north,legend columns=-1},
            symbolic x coords={SG-Single,
SG-1shard,
SG-12shard,
Neo4J,
Neo4J-rep2,
Neo4J-rep3,
JanusGraph,
JanusGraph-4shard},
            xtick=data,
            x tick label style = {font = \scriptsize, text width = 1.4cm, align = center, rotate = 70, anchor = north east},
            %nodes near coords,
            %nodes near coords align={vertical},
            ymin=0,ymax=450000,
            ylabel={Insertion Rate (Vertices/sec)},
            ylabel style={yshift=-0.5cm}
        ]
        \addplot table[x=name,y=rate]{data/vertex/bar_vertices.txt};
        %\addplot table[x=interval,y=carD]{\mydata};
        %\addplot table[x=interval,y=carR]{\mydata};
        %\legend{Trips, Distance, Energy}
    \end{axis}
\end{tikzpicture}

%\centerline{\includegraphics[scale=0.6]{experiments/all_add_vertex.eps}}
\centerline{(a) Average throughput}

\end{minipage}
 \ \hspace*{0.4cm}
 \begin{minipage}[b]{0.48\textwidth}

\begin{tikzpicture}
\begin{groupplot}[
    group style={
        group name=my fancy plots,
        group size=1 by 2,
        xticklabels at=edge bottom,
        vertical sep=0pt
    },
    width=8.5cm,
    xmin=0, xmax=50
]

\nextgroupplot[ymin=350000,ymax=450000,
               ytick={400000, 450000},
               axis x line=top, 
               axis y discontinuity=parallel,
               %ytick scale label code/.code={},
               height=9em]
\addplot [mark=o, color=blue]table [x=batch, y=rate]{data/vertex/plot_dist12.txt};     

\nextgroupplot[ymin=0,ymax=150000,
               ytick={0, 50000, 100000, 150000},
               axis x line=bottom,
               ytick scale label code/.code={},
               ylabel={Insertion Rate (Vertices/sec)},
               ylabel style={xshift=0.8cm, yshift=-0.5cm},
                legend style={
                    font=\tiny,
                    row sep=0,
                    at={(0.5,-0.25)},
                    anchor=north,
                    legend columns=3,
                    %/tikz/every even row/.append style={row sep=.1em}
                        },
               height=11em]
           
                \addplot [mark=diamond,color=BurntOrange]  table  [x=batch, y=rate]{data/vertex/plot_single.txt};
                \addplot [mark=x, color=olivegreen]  table [x=batch, y=rate]{data/vertex/plot_dist1.txt};
               \addplot [mark=o, color=blue]  table [x=batch, y=rate]{data/vertex/dist12_dummy.txt};    
                \addplot [mark=square, color=gray] table [x=batch, y=rate]{data/vertex/plot_neo4j.txt};
                \addplot [mark=+, color=red] table [x=batch, y=rate]{data/vertex/plot_janus.txt};     
                \legend{SG-Single, SG-1shard, SG-12shard, Neo4J, JanusGraph}
\end{groupplot}
\end{tikzpicture}

\centerline{(b) Vertex insertion rates for consecutive blocks of 1 million vertices}
 \end{minipage}
 \caption{\label{fig:vertex} Vertex Insertion Rates.}
\end{figure*}

\label{sec:experiments}

In this section we evaluate the performance of core graph database
operations by adding vertices, adding edges and performing simple Breadth
First Search (BFS) queries. We look at the scalability of the
distributed graph database, analyze sources of overhead when comparing
the distributed database versus the single node.
We also compare SystemG
against other existing graph databases: Neo4J Enterprise~\cite{neo4j2013}, version
3.2.1, and JanusGraph~\cite{janusgraph} 0.1.1, a successor of Titan~\cite{titan} using Apache Cassandra 3.7~\cite{cassandra} 
as its backend. 
We faithfully did our best to optimize the performance of Neo4J and JanusGraph by carefully writing the application scripts, choosing the right batching and size of clusters. 

The experiments were performed using a cluster with five IBM S824L~\cite{s824l} machines. 
The configuration is displayed in Table~\ref{table:parameters}.
Each had 4 numa nodes, 24 POWER8E cores with smt8 at 3.3GHz, 2TB memory. 
One machine was used to run query manager and others were used for shards, which we call the servers. 
The machine for query manager connects to an external IBM
FlashSystem 840 Enterprise SSD storage~\cite{flash840} for loading input data files. 
We will refer to this machine as the frontend.
It is used to run clients in the distributed
experiments for our SystemG, Neo4J, and JanusGraph.
The machines are connected to each other through a 100 Gbit switch~\cite{mellanox-sb7790}.

\subsection{Vertex Addition}

In this section we analyze the performance of adding vertices to the
database. We analyze several different configurations measuring throughput
for adding vertices and the results are included in Figure
\ref{fig:vertex}. The input data file used is derived from Twitter 
datasets~\cite{twitter} and it contains 58,269,482 vertices, each
vertex with its external identifier and two properties.  We assume the
user will perform look-ups based on this identifiers so they need to
be indexed. While both SystemG and JanusGraph provide by default an index for external identifiers, for Neo4J we had to explicitly create
an index on a property ``ID'' and treat the external id as an indexed
property.

 The first configuration analyzed is the single node
version of the SystemG using the C++ API (\emph{SG-Single}), 
we committed every 1 million operations making sure data was completely
persisted to disk. In our model this is achieved using {\tt
graph::tx\_commit(RDWR)}.
The next configuration considered was distributed graph database using
one shard, a query manager node and a Firehose node (\emph{SG-1shard}). The database
shard and query manager were run on a server and the Firehose was run
on the frontend. The protocol used for adding vertices in the
distributed mode was to read from the input file entries, aggregate up
to 100,000 entries and then send them as one command to the shard to
insert. A third scenario is the distributed database with 12 shards (\emph{SG-12shard}),
3 in each of the 4 servers. We choose this number because we observed through
experimentation that ingesting data scales up to three\textasciitilde six shards per
node. After this point adding more shards will not benefit much as the
available resource becomes saturated (see Section~\ref{sec:scale}). When using 12 shards the Firehose will
batch 1.2 million vertices (12*100K) in 12 queues and then 12 threads
will contact shards to insert queued vertices in parallel. Thus we
expect very good scaling when adding vertices using multiple shards.
The transactions are generated and maintained on the frontend
machine on the cluster, and committed evenly and asynchronously using one
thread per machine to the server.

%while in Neo4J we used {\tt
%tx.success(); tx.close();}. 

For Neo4J, we implemented the embedded Java API for a single machine (\emph{Neo4J}). 
Also, we implemented multi-node configurations with replica nodes using the  high-availability for 1 server, 3 instances (\emph{Neo4J-3rep}) and 2 servers, 2 instances (\emph{Neo4J-2rep}) with HTTP APIs. 
Even though evaluating replicas with distributed databases is not at all a fair comparison, we include them to see whether they are beneficial for performances since Neo4J does not support a distributed mode. 
After some testing, we found that the 
transactional Cypher HTTP API for Neo4J significantly degrades in 
performance for large queries; thus, we committed using batches of size 
20,000 instead of 100,000.

\begin{figure*}[t]
\begin{minipage}[b]{0.48\textwidth}
%\centerline{\includegraphics[scale=0.6]{experiments/methods.eps}}

\begin{tikzpicture} 
\begin{axis}[ 
    ybar,
    ymode=log,
    width = \textwidth,
    bar width=.2cm,
    height=14em,
    xtick={1,...,20},
    x tick label style = {font = \small, text width = 1.5cm, align = right, rotate = 90, anchor = north east, yshift=0.7em},
    xticklabels from table = {data/edge/SG.txt}{name},
    xtick align=inside,
    every axis x label/.style={at={(ticklabel cs:0.5)},anchor=near ticklabel},
    ymin=1000,ymax=500000,
            ylabel={Insertion Rate (Edges/sec)},
    every axis y label/.style={at={(ticklabel cs:0.5)},rotate=90,anchor=near ticklabel},
             legend style={at={(0,1)},
                anchor=north west},
]

\addplot[ybar, bar shift=0pt, fill=cyan!70,
    discard if not={group}{Sync},
    ] table [ 
    x=index, 
    y=rate
] {data/edge/SG.txt} ;

\addplot[ybar, bar shift=0pt, fill=orange!70,
    discard if not={group}{Async},
   ] table [ 
    x=index, 
    y=rate
] {data/edge/SG.txt} ;

\addplot[ybar, bar shift=0pt, fill=gray!70,
    discard if not={group}{FH},
   ] table [ 
    x=index, 
    y=rate
] {data/edge/SG.txt} ;
\legend{Sync, Async, Firehose}

\end{axis} 
\end{tikzpicture} 
\centerline{(a) Add Edge rates: Synchronous, Asynchronous and Firehose}
\end{minipage}
 \ \hspace*{0.4cm}
 \begin{minipage}[b]{0.48\textwidth}
%\centerline{\includegraphics[scale=0.6]{experiments/add_edge.eps}}
\begin{tikzpicture}
    \begin{axis}[
            ybar,
            bar width=.5cm,
            width=\textwidth,
            height=14em,
            %legend style={at={(0.5,1)},
            %    anchor=north,Neo4 columns=-1},
            symbolic x coords={SG-Single,
SG-1shard,
SG-12shard,
Neo4J,
Neo4J-rep2,
Neo4J-rep3,
JanusGraph,
JanusGraph-4shard},
            xtick=data,
            x tick label style = {font = \scriptsize, text width = 1.4cm, align = center, rotate = 70, anchor = north east},
            %nodes near coords,
            %nodes near coords align={vertical},
            ymin=0,ymax=150000,
            ylabel={Insertion Rate (Edgees/sec)},
            ylabel style={yshift=-0.5cm}
        ]
        \addplot table[x=name,y=rate]{data/edge/bar_edges.txt};
        %\addplot table[x=interval,y=carD]{\mydata};
        %\addplot table[x=interval,y=carR]{\mydata};
        %\legend{Trips, Distance, Energy}
    \end{axis}
\end{tikzpicture}
\centerline{(b) Average Edge Insertion Rates}
 \end{minipage}
 \caption{\label{fig:edge} Edge Insertion Rates.}
\end{figure*}
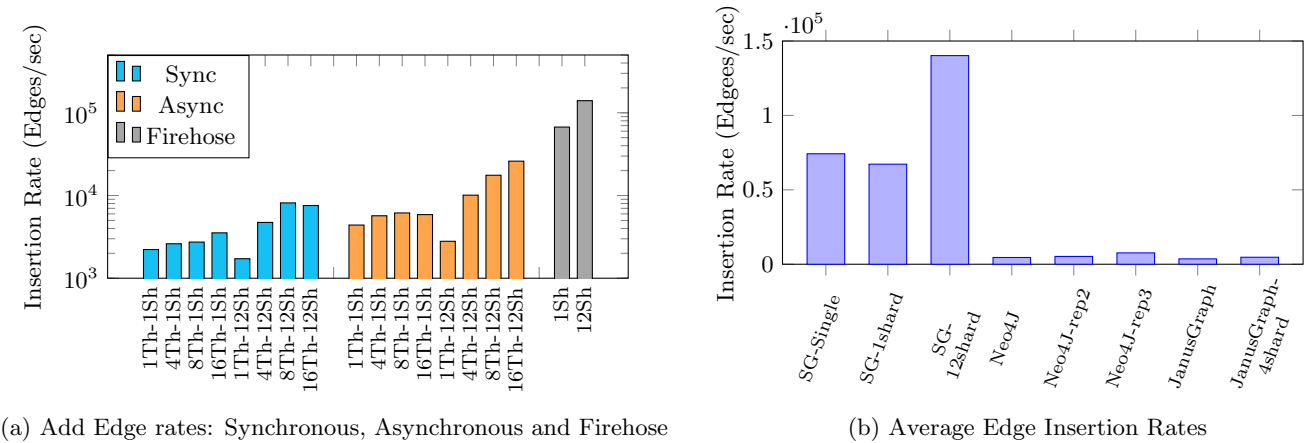

%we read from the input file, aggregate up to 20,000 transactions, and commit as a batch. 

%In order to assess the degree of scalability for this database, we also implemented a high-availability
%cluster of three instances on a single machine. 

For JanusGraph, we tested by ingesting vertices using the TinkerPop Core
API frontend and using Apache Cassandra 3.7 as a storage backend, with 
recommended settings for both JanusGraph and Cassandra from the official
documentation. The first scenario implemented uses a single instance of
Cassandra (\emph{JanusGraph}).
We also tested JanusGraph with a distributed Cassandra
storage backend across four machines in the cluster (\emph{JanusGraph-4shard}).

%The second scenario uses the storage.batch-loading setting for JanusGraph,
%which disables internal consistency checking and assumes data consistency.
%For the sake of fairness, all other parameters are set identically to the
%first scenario. Finally, we tested JanusGraph with a distributed Cassandra
%storage backend across four machines in the cluster. In this case, batches
%of 200,000 are used, and transactions are made to an arbitrary one of the
%four storage instances.
%\wvijit{include janusgraph results in fig vertex}

Figure \ref{fig:vertex} (a) shows the average insertion rate for
ingesting the whole input file expressed as vertices per second. The
single node SystemG averages 102,455 vertices per second, with one
shard 98,979 and distributed with 12 shard 415,173 vertices per second.
The distributed with one shard incurs a small throughput degradation due
to the additional processing and communication overhead of query manager but this loss is recouped when
scaling the number of shards, up to 4.2$\times$ when using 4 machines with 12 shards. Other important factors affecting the
performance of all modifying graph operations are the granularity of the
transactions, how many threads are ingesting at the same time and the storage
used, all of which could not be separately evaluated due to lack of time
and resources. For comparison, Neo4J achieved single node 93,235, two
nodes 69,573, and three nodes 60,797 vertices per second. 
%For adding vertices, the slowdown for Neo4J in multiple nodes is expected since all vertices have to be broadcasted to other nodes since they are replicas.
For Neo4J, the throughput achieved for vertex
ingestion is not strongly correlated with either the size of the cluster
or to the number of separate instances in deployment since they are replicas and the added vertices have to broadcasted.
Also there is a significant overhead associated with the Cypher HTTP API that is
unavoidable. % for configurations involving remote server instances. 
JanusGraph achieved single node 11,477 vertices per second and four node distributed 11,007 vertices per second, showing low performance among all databases in comparison. 
Overall, Single node SystemG was able to achieve about 9.9\% speedup over Neo4J and 8.9$\times$ that of JanusGraph. 
When scaled with 12 shards, the speedup of SystemG was 4.2$\times$ and 36.2$\times$ over the best configurations for Neo4J and JanusGraph, respectively.

In Figure \ref{fig:vertex} (b) we show the ingest rates for adding
vertices. We plot the
statistics every 1 Million operations.
The performances are relatively stable for all single-shard databases.
%except for distributed with 12 shards where statistics are printed every 1.2M operations. 
%We observe good scaling for the single node graph database. 
There are slight slowdowns as the
underlying btree structure grows inside the LMDB used for backend storage.
For SG-12shard, it exhibits performance boxed above 400,000 vertices per second, which is superior over all others. 
The small fluctuation accounts for the network traffic congestion in communication from the query manager machine to the servers.
%e is using a btree based structure we should
%see a slow decrease in performance as the btree grows. 
%For this
%experiment Neo4J is using all available computing resources, the
%single node and distributed one shard are using a single thread, while
%distributed 12 shards is using six threads per shard.
%\jinho{janus}

\subsection {Edge Addition}

In this section we analyze the performance of adding edges to the
database. The semantic of adding an edge is as described in Figure
\ref{fig:ae}. The data input considered is a set of edges
derived from bitcoin transactions and it contains three columns:
source, target, and weight which is a double value
A first experiment shown in Figure \ref{fig:edge} (a), was to evaluate
the performance of the three methods of adding edges as described in
section \ref{sec:methods}. First method called synchronous (SYNC) will
follow through the eight communication steps sequence described in Section
\ref{sec:sync}. The second method we evaluate is the asynchronous
version (ASYNC) described in Section \ref{sec:async}. We performed the
four communication steps required to confirm to the user that the method finished
all its steps on the various shards. Finally we evaluate the batched
Firehose method as shown in Section \ref{sec:firehose}. 
For SYNC and ASYNC we execute with 1 shard and 12 shards with four machines.
%scenarios using a client on the frontend machine
%and the distributed database with one and six shards running in one
%server. 
Then we vary the number of threads requesting the add edge methods at the query manager side from one to 16. 
In the legend, for SYNC and ASYNC methods, $n$Th-$m$Sh is used to represent the configuration with $n$ requesting threads with $m$ shards. 
%We call the add edge methods from one thread running on
%frontend against the server running with one shard (Sync-1Th-1Sh and
%Async-1Th-1Sh), next one thread against the database with six shards
%(Sync-1Th-6Sh and Async-1Th-6Sh). Next we call methods from the client
%in a concurrent fashion using either 4 or 8 threads against the six
%shards distributed database. 
While concurrent write transactions against
a single node database or shards does not add much benefit due to
serialization of transactions, in a multishard environment we do
expect improved throughput from running multiple client threads to
ingest data. When inserting concurrently with P threads the input file
was split in P chunks and each thread inserted one chunk, thereby avoiding duplicate edge insertions.

In Figure \ref{fig:edge} (a), we see the SYNC methods achieving 2,223 edges/sec in Sync-1Th-1Sh to 7,571 edges/sec in Sync-16Th-12Sh. 
%Sync-1Th-1Sh,
%Sync-1Th-6Sh, Sync-4Th-6Sh, Sync-8Th-6Sh achieving 2,936, 2,801, 9,832
%and 13,917 edges/sec respectively. 
For the experiment in this section
we only ingest 10 million bitcoin edges. When going from one shard to multiple
shards there is a small overhead for maintaining additional
state. However this overhead is compensated for by running inserts
concurrently. In this particular case we obtain a 3.4$\times$ speedup by
ingesting data with 16 threads on 12 shards versus one thread and one
shard.

For ASYNC methods, the performance ranges from 4,393 edges/sec in Async-1Th-1Sh to 26,025 edges/sec in Async-16Th-12Sh. 
%Async-1Th-1Sh, Async-1Th-6Sh, Async-4Th-6Sh, Async-8Th-6Sh we
%obtained 7,782, 5,262, 18,730 and 30,599 edges/second
%respectively. 
The reduced number of messages exchanged and the use of
asynchronous methods leads to a good performance improvement compared
to the synchronous version. For eight threads and six shards the
speedup of the asynchronous version is 5.9$\times$ relative to one thread and
one shard. 

The Firehose batched version implements the more
complicated protocol described in Section
\ref{sec:firehose} but it has the advantage of committing methods in
bulk thus increasing the granularity of transactions and improve the
overall throughput. In Figure \ref{fig:edge} (a) the Firehose will
commit in batches of 100K edges when using one shard and batches of
12$\times$100K when using 12 shards. Overall, for the two cases we obtain
67,249 and 140,208 edges per second for the single shard, and 12 shards, respectively.

On the single shard configurations, speedup of ASYNC was 2.0$\times$ over SYNC, and that of Firehose was 30.3$\times$. With 12 shards, ASYNC was 3.4$\times$ over SYNC, and Firehose was 18.5$\times$ faster than SYNC method.
While the Firehose is definitely the solution for ingesting large
amount of edges it may not be appropriate in every situation. If an
add edge method can not be batched and it needs to be visible right away
the SYNC or ASYNC version will have to be used.

In Figure \ref{fig:edge} (b) we show the overall ingestion rates when
ingesting 10 million edges from the bitcoin dataset. We commit every 100k
edges for SystemG and JanusGraph tests except distributed with 12 shards
where we commit every 12x100k edges. For Neo4J, we committed every 20k
edges for each test due to the same issue of performance degradation for
large Cypher HTTP API transactions. We tested SystemG with distributed one
shard and distributed 12 shard, Neo4J in a highly available configuration with one-, two-, and three-
instance clusters, and JanusGraph with one, four Cassandra instance.
The JanusGraph benchmark was written in Java
using TinkerPop Core API version 2.5 and run in embedded mode
(e.g., we did not run JanusGraph as a server). We used separate servers
for running Cassandra.
%, and did not enable batch loading due to possible
%loss of consistency due to concurrent transactions. 

The single node SystemG achieves 74,247 edges/second, 
and 
distributed one shard gets 67,249 edges/second with a small expected drop from the single node SystemG.
Distributed 12 shards shows
140,208 edges/second, which scaled well despite the multiple round-trip communications per transaction. 
Performances of Neo4J are one
instance 4,530 edges/second, two instance 5,280, and three instance 7,634.
While Neo4J replica instances theoretically does not improve the addition of edges itself, checking for vertices can be done locally because all instances are exact same clones.
That results in a slight speedup when multiple nodes are used, but was not very significant. 
JanusGraph achieved 3,614 edges/second for 1 Cassandra instance, and 3240 edges/second for 4 Cassandra instances on average during the run, and logs
indicate a marked drop in performance as more vertices were inserted into
the database. 

Performance for SystemG was significantly higher, 
The speedup of SystemG on single node is 16.4$\times$ over Neo4J and 20.5$\times$ over JanusGraph. With multiple shards, SystemG was 18.3$\times$ faster than Neo4J, and 38.8$\times$ over JanusGraph on their best configurations.

\subsection{Query analysis}

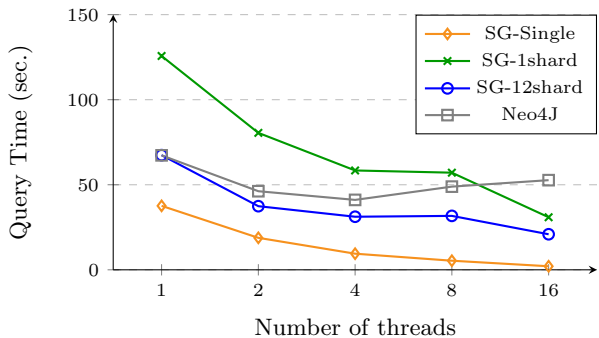
\begin{figure}[t]
%\begin{minipage}[b]{0.48\textwidth}
%\centerline{\includegraphics[scale=0.6]{experiments/query.eps}}

 \begin{tikzpicture}
    \begin{axis}[
                    width=.95\linewidth,
                    height=.59\linewidth,
                    axis lines = left,              
                    xlabel={Number of threads},
                    xmin=0.5, xmax=5.5,
                    ymin=0, ymax=150,
                    xtick = data,
                    xticklabels={1, 2, 4, 8, 16},
                    ymajorgrids=true,
                    grid style=dashed,
                    ylabel={Query Time (sec.)},
                    tick label style={font=\scriptsize},
                    legend style={font=\scriptsize, at={(1,1)}},
                   ]
                   
                \addplot [mark=diamond,color=BurntOrange, thick]  table  [x=index, y=time]{data/bfs/single.txt};
                \addplot [mark=x, color=olivegreen, thick]  table [x=index, y=time]{data/bfs/dist1.txt};
               \addplot [mark=o, color=blue, thick]  table [x=index, y=time]{data/bfs/dist12.txt};    
                \addplot [mark=square, color=gray, thick] table [x=index, y=time]{data/bfs/neo4j.txt};
                 
                \legend{SG-Single, SG-1shard, SG-12shard, Neo4J}
    \end{axis}
\end{tikzpicture}
\centerline{(a) BFS query time; 100K queries } 
%\end{minipage}
% \ \hspace*{0.4cm}
% \begin{minipage}[b]{0.48\textwidth}
%\centerline{\includegraphics[scale=0.6]{experiments/add_edge.eps}}
%\centerline{(b) placeholder}
% \end{minipage}
 \caption{\label{fig:query} BFS query analysis}
\end{figure}

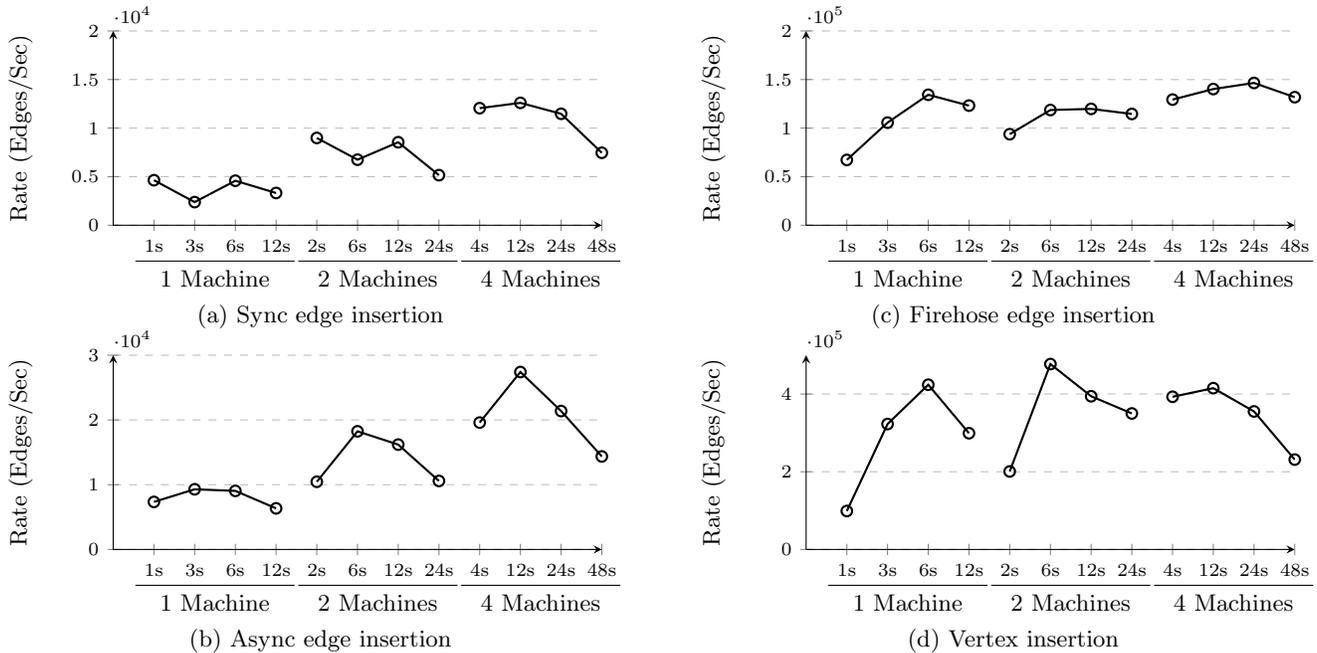
\begin{figure*}[t]

\pgfplotstableread{
1 1
2 1
3 1
4 1
5 2
6 2
7 2
8 2
9 3
10 3
11 3 
12 3
}\datatable

 \begin{minipage}[b]{0.48\textwidth}
%\centerline{\includegraphics[scale=0.6]{experiments/add_edge.eps}}
 \begin{tikzpicture}
    \begin{axis}[
                    width=.95\linewidth,
                    height=.49\linewidth,
                    axis lines = left,              
                    xlabel style={yshift=-0.5cm},
                    xmin=0, xmax=12,
                    ymin=0, ymax=20000,
                    xtick = {1,2,3,4,5,...,12},
                    xticklabels={1s, 3s, 6s, 12s, 2s, 6s, 12s, 24s, 4s, 12s, 24s, 48s},
                    ymajorgrids=true,
                    grid style=dashed,
                    ylabel={Rate (Edges/Sec)},
                %    y label style={at={(axis description cs:.1,.5)},anchor=south},
                    %ytick = {0,1,2,3,4},
                    %axis line style = ultra thick,
                    %every axis plot/.append style={thick}
                    tick label style={font=\scriptsize},
                    legend style={font=\scriptsize, at={(1,1)}},
                   draw group line={[index]1}{1}{1 Machine}{-3.5ex}{7pt},
    draw group line={[index]1}{2}{2 Machines}{-3.5ex}{7pt},
    draw group line={[index]1}{3}{4 Machines}{-3.5ex}{7pt},
                   ]
                   
                \addplot [mark=o,color=black, thick]  table  [x=index, y=rate]{data/scal_sync/1m.txt};
                \addplot [mark=o,color=black, thick]  table [x=index, y=rate]{data/scal_sync/2m.txt};
               \addplot [mark=o,color=black, thick]  table [x=index, y=rate]{data/scal_sync/4m.txt};    
                 
                %\legend{SG-Single, SG-1shard, SG-12shard, neo4j}
    \end{axis}
\end{tikzpicture}
\centerline{(a) Sync edge insertion}
 \begin{tikzpicture}
    \begin{axis}[
                    width=.95\linewidth,
                    height=.49\linewidth,
                    axis lines = left,              
                    xlabel style={yshift=-0.5cm},
                    xmin=0, xmax=12,
                    ymin=0, ymax=30000,
                    xtick = {1,2,3,4,5,...,12},
                    xticklabels={1s, 3s, 6s, 12s, 2s, 6s, 12s, 24s, 4s, 12s, 24s, 48s},
                    ymajorgrids=true,
                    grid style=dashed,
                    ylabel={Rate (Edges/Sec)},
                %    y label style={at={(axis description cs:.1,.5)},anchor=south},
                    %ytick = {0,1,2,3,4},
                    %axis line style = ultra thick,
                    %every axis plot/.append style={thick}
                    tick label style={font=\scriptsize},
                    legend style={font=\scriptsize, at={(1,1)}},
                   draw group line={[index]1}{1}{1 Machine}{-3.5ex}{7pt},
    draw group line={[index]1}{2}{2 Machines}{-3.5ex}{7pt},
    draw group line={[index]1}{3}{4 Machines}{-3.5ex}{7pt},
                   ]
                   
                \addplot [mark=o,color=black, thick]  table  [x=index, y=rate]{data/scal_async/1m.txt};
                \addplot [mark=o,color=black, thick]  table [x=index, y=rate]{data/scal_async/2m.txt};
               \addplot [mark=o,color=black, thick]  table [x=index, y=rate]{data/scal_async/4m.txt};    
                 
                %\legend{SG-Single, SG-1shard, SG-12shard, neo4j}
    \end{axis}
\end{tikzpicture}
\centerline{(b) Async edge insertion}
 \end{minipage}
  \ \hspace*{0.4cm}
 \begin{minipage}[b]{0.48\textwidth}
%\centerline{\includegraphics[scale=0.6]{experiments/query.eps}}

 \begin{tikzpicture}
    \begin{axis}[
                    width=.95\linewidth,
                    height=.49\linewidth,
                    axis lines = left,              
                    xlabel style={yshift=-0.5cm},
                    xmin=0, xmax=12,
                    ymin=0, ymax=200000,
                    xtick = {1,2,3,4,5,...,12},
                    xticklabels={1s, 3s, 6s, 12s, 2s, 6s, 12s, 24s, 4s, 12s, 24s, 48s},
                    ymajorgrids=true,
                    grid style=dashed,
                    ylabel={Rate (Edges/Sec)},
                %    y label style={at={(axis description cs:.1,.5)},anchor=south},
                    %ytick = {0,1,2,3,4},
                    %axis line style = ultra thick,
                    %every axis plot/.append style={thick}
                    tick label style={font=\scriptsize},
                    legend style={font=\scriptsize, at={(1,1)}},
                   draw group line={[index]1}{1}{1 Machine}{-3.5ex}{7pt},
    draw group line={[index]1}{2}{2 Machines}{-3.5ex}{7pt},
    draw group line={[index]1}{3}{4 Machines}{-3.5ex}{7pt},
                   ]
                   
                \addplot [mark=o,color=black, thick]  table  [x=index, y=rate]{data/scal_fe/1m.txt};
                \addplot [mark=o,color=black, thick]  table [x=index, y=rate]{data/scal_fe/2m.txt};
               \addplot [mark=o,color=black, thick]  table [x=index, y=rate]{data/scal_fe/4m.txt};    
                 
                %\legend{SG-Single, SG-1shard, SG-12shard, neo4j}
    \end{axis}
\end{tikzpicture}
\centerline{(c) Firehose edge insertion}
 \begin{tikzpicture}
    \begin{axis}[
                    width=.95\linewidth,
                    height=.49\linewidth,
                    axis lines = left,              
                    xlabel style={yshift=-0.5cm},
                    xmin=0, xmax=12,
                    ymin=0, ymax=500000,
                    xtick = {1,2,3,4,5,...,12},
                    xticklabels={1s, 3s, 6s, 12s, 2s, 6s, 12s, 24s, 4s, 12s, 24s, 48s},
                    ymajorgrids=true,
                    grid style=dashed,
                    ylabel={Rate (Edges/Sec)},
                %    y label style={at={(axis description cs:.1,.5)},anchor=south},
                    %ytick = {0,1,2,3,4},
                    %axis line style = ultra thick,
                    %every axis plot/.append style={thick}
                    tick label style={font=\scriptsize},
                    legend style={font=\scriptsize, at={(1,1)}},
                   draw group line={[index]1}{1}{1 Machine}{-3.5ex}{7pt},
    draw group line={[index]1}{2}{2 Machines}{-3.5ex}{7pt},
    draw group line={[index]1}{3}{4 Machines}{-3.5ex}{7pt},
                   ]
                   
                \addplot [mark=o,color=black, thick]  table  [x=index, y=rate]{data/scal_v/1m.txt};
                \addplot [mark=o,color=black, thick]  table [x=index, y=rate]{data/scal_v/2m.txt};
               \addplot [mark=o,color=black, thick]  table [x=index, y=rate]{data/scal_v/4m.txt};    
                 
                %\legend{SG-Single, SG-1shard, SG-12shard, neo4j}
    \end{axis}
\end{tikzpicture}
\centerline{(d) Vertex insertion} 
\end{minipage}
 \caption{\label{fig:scale} Performance scalability of SystemG.}
\end{figure*}

In this section we analyze the performance of a simple BFS query
against the data in various databases. While the distributed
database is often very good at ingesting data and improving
the query throughput, the tradeoff is that it often increases the
execution time or latency for a single graph query. The reason for this
is the fact that for most graph queries which takes the form of a
graph traversal the data will be collected from multiple shards in an
iterative process. The algorithm that we evaluate in this section was
introduced in Section \ref{sec:bfs}, Figure \ref{fig:bfs}. We
basically perform BFS from a starting point and we stop the traversal
after a given depth. This pattern is the backbone for various graph
searches. We bound the depth of the traversal as the number of edges
traversed may grow exponentially to the point where it can very quickly reach the
whole graph~\cite{bfs_gpu}. While this may be required in certain situations in
practice, various application domain may impose a bound on the search
space to guarantee certain performance bounds for a particular
query. 
For example in the financial domain when deciding if a bank
transaction is fraudulent or not the database often has only a few hundred
milliseconds to perform the analysis. While a complete search is
desirable, in such short amount of time only a subset of data can be
practically analyzed. 
For all BFS queries discussed in this section, we used the Higgs Twitter dataset~\cite{higgs} which depicts a sparse graph of 14,855,842 directed edges
in edgelist format, without property values.

For all BFS queries analyzed in this section we stop after four levels
deep and we return as part of the query all unique vertices accessed
by the traversal. For Neo4J we implemented the algorithm using the
native Java API and we run the experiment in embedded mode similar to
the SystemG single node. For the distributed case we evaluated one and
twelve shards case, similar to previous experiments. While running distributed there is a query manager
on one of the servers together with the shards of the database. Each
shard will employ one master thread and the scheduler will use another
four threads for executing incoming RPC with requests for data. In
Figure \ref{fig:query} we show the result for performing 100K queries
using 100K different start vertices. We used the same starting
vertices, identified by external IDs, for all solutions considered
and we verified that all solutions return the same number of edges
traversed for each starting vertex. We considered a
variable number of threads and when running with P threads the set of
100K start vertices was split in P blocks each thread performing the
queries in one block. 
This is not a perfect way to split the work across
threads because each query will be different, causing different
amounts of work and possible leading to some unbalance between
threads. However we don't refine this in this work.

As we can see the SystemG single node graph database has the best
performance overall finishing all 100K queries in 37.6 seconds using one
thread. The results returned for individual traversals vary in size
from one edge to a couple of thousand edges for the Higgs Twitter edge set
considered. When using 16 concurrent threads the execution time for
all queries reduces to 3.5 seconds for a 10.7$\times$ speedup. The speedup is
not linear even though we mainly do read only operations. This is because
all threads are running in the same memory address space and each query
will allocate memory for internal queues and result vectors causing
contention on the memory allocator. Neo4J scales from 67.3 seconds when
using one thread to 41.1 seconds when using 4 threads and goes up after that. 
For Neo4J
running more user threads impacts the database threads when running in
embedded mode as they will all need to be scheduled on the same
resources. %However the machine had 60 cores and 120 hardware threads.
For distributed one shard we pay the communication and one extra
indirection overhead. In this case the client which runs on the
frontend sends the request to the query manager which will execute the
traversal for the client. The traversal involves a number of
communications with the shards and when finished, the query manager
will send the results back to the client. For the case when using one
shard we essentially measure the overhead of the distributed
implementation. We see the total execution time to be 125.7 seconds when
using one client thread and 30.9 seconds when using 16 client threads
for a speedup of 4.1$\times$. 
%Additional experiments we performed showed
%better speedups when considering 1 million queries. For the single
%node SystemG 1 Million queries finished in 411 seconds with one thread
%and 55 seconds with 32 threads for a speedup of 2.20. 
For distributed
with 12 shards we scale from 67.4 seconds for one thread to 20.9 seconds
with 16 threads. With 12 shards, it still pays the communication delay, but the running time decreases due to parallel query processing for retrieving neighbor edges.
Thus, we conclude that while there is a significant overhead of distributed querying compared to a single shard this is compensated for as we run
more client threads, and more shards. 
The speedup of SystemG over Neo4J was 19.7$\times$ with the single node.
Even with the overhead of distributed settings, SystemG obtained 1.3$\times$ speedup with 1 shard and 2.0$\times$ speedup with 12 shards.

We performed the query benchmark for JanusGraph as well using essentially
the same code as for Neo4J with only few lines changed corresponding
to the differences in API. The execution times however were much
longer compared to Neo4J and SystemG. We estimated it will take
hours to finish the initial set of 100K queries and decided not to
complete the full test. This correlates with other results discussed in
the literature~\cite{Beis2015}.
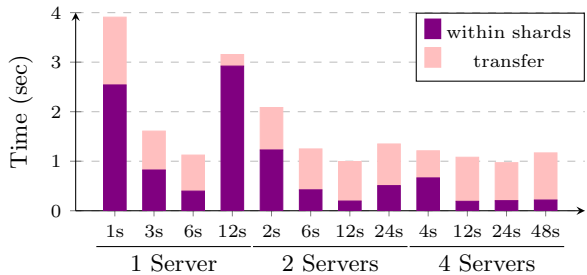
\begin{figure}
\pgfplotstableread{
1 1
2 1
3 1
4 1
5 2
6 2
7 2
8 2
9 3
10 3
11 3 
12 3
}\datatable
\begin{tikzpicture}
    \begin{axis}[
                    ybar stacked,
                    width=.99\linewidth,
                    height=.5\linewidth,
                    bar width=0.3cm,
                    axis lines = left,              
                    ylabel style={yshift=-0.5cm},
                    xmin=0, xmax=13,
                    ymin=0, ymax=4,
                    xtick = {1,2,3,4,5,...,12},
                    xticklabels={1s, 3s, 6s, 12s, 2s, 6s, 12s, 24s, 4s, 12s, 24s, 48s},
                    ymajorgrids=true,
                    grid style=dashed,
                    ylabel={Time (sec)},
                %    y label style={at={(axis description cs:.1,.5)},anchor=south},
                    %ytick = {0,1,2,3,4},
                    %axis line style = ultra thick,
                    %every axis plot/.append style={thick}
                    tick label style={font=\scriptsize},
                    legend style={font=\scriptsize, at={(1,1)}},
                   draw group line={[index]1}{1}{1 Server}{-3.5ex}{7pt},
    draw group line={[index]1}{2}{2 Servers}{-3.5ex}{7pt},
    draw group line={[index]1}{3}{4 Servers}{-3.5ex}{7pt},
                   ]
                   
                \addplot [ybar, fill = violet, color = violet]  table  [x=index, y=proc]{data/breakdown/break.txt};
                \addplot [ybar, fill = violet, color = pink]  table [x=index, y=comm]{data/breakdown/break.txt};
                 
                \legend{within\ shards, transfer}
    \end{axis}
\end{tikzpicture}
 \caption{\label{fig:breakdown} Timing breakdown on vertex addition.}
\end{figure}

\subsection{Scalability}
\label{sec:scale}
\sloppypar
In this section, we show how the performance of SystemG scales with the number of Servers and shards.
In Figure~\ref{fig:scale} (a), (b), and (c), the performance change of adding edges is shown, with the three methods of adding edges (SYNC, ASYNC, and Firehose).
For the configuration, we used the combinations of 1, 2, 4 Servers and 1, 3, 6, 12 shards per server.
We used 16 threads that make queries for SYNC and ASYNC edge additions.  
For SYNC and ASYNC algorithms, the performance scales well as the number of shards go up. 
The trend is more clear on ASYNC algorithms, since the number of communications between the query manager and the servers is much less.
However, it can be seen that the performance starts to drop at around 3\textasciitilde 6 shards per server. 
Each SystemG shard uses 6 threads internally, and if there are more threads in total than the number of cores (24 cores, which fits around 4 shards per server), context switching and scheduling overhead causes adversarial effects to the database even though the number of hardware threads are much more since the cores are running with smt8.  

With Firehose however, the performance saturates quicker. 
The addition rate goes up steeply with one server until 6 shards, but the benefit from adding more servers is marginal.
A similar trend can be seen in Figure~\ref{fig:scale} (d), where we show the performance change of adding vertices with Firehose.
It almost scales proportional to the aggregate number of shards when there are fewer number of shards, because for adding vertices, there is no round-trip or inter-shard communications. 
However, it saturates when there are more than around 6 shards in total. 
The saturation in using Firehose comes from the network bandwidth bottleneck at the query manager node, for sending the raw vertices/edges data to the shards.
Figure~\ref{fig:breakdown} shows the average timing breakdown sampled while adding a few batches of 300,000 vertices to the shards.
As the performance saturates, there is very little time spent in the shards for actual processing down to far less than a second, and most of the time is spent on the network. On the other hand, the time spent within the shards usually decreases inversely proportional to the number of shards, because the same amount of batched vertices are split among the shards. However, increase of the time spent in the shards is observed for 1 server-12 shards, 2 servers-24 shards, and 4 servers-48 shards cases.
It indicates that context switching and scheduling overhead is becoming severe due to too many threads.
Since the ratio do not change much when we changed the size of the batch, it is a bandwidth problem, not a latency problem. 
We plan to reduce the network overhead by optimizing the messaging protocol as a future work.

\section{Conclusion}
\label{sec:conclusion}

In this paper, we presented SystemG, a single node graph database and its extension as a distributed graph
database designed for ingesting and querying big relational data to meet the industry needs. We
have shown the high level design and the main supporting modules like
an efficient task based runtime system and RPC driven communication to
allow for high throughput insertions as well as many parallel low latency queries. We
introduced two novel concepts for a graph database: Firehose
used for fast batched insertions and the Query Manager that will
perform complex queries for client before returning results. We also
presented two novel methods for adding edges in the context of
undirected or directed with predecessors graphs. We evaluated the
performance of both the single node and distributed graph database and
provided for comparison performance numbers for two well established
graph databases. Our graph solutions provide significantly better performances
overall at the cost of a more relaxed consistency model that need to
be compensated for by the programmer. As a future work, we are working on adding
some very important features such as fault tolerance, distributed
asynchronous queries and eventually distributed transaction support.

\bibliographystyle{abbrv}
\bibliography{graph_short}  % sigproc.bib is the name of the Bibliography in this ca

% that's all folks
\end{document}